\documentclass{article}
\usepackage{graphicx} 

\title{Optimizing for Rotisserie Fantasy Basketball}
\author{Zach Rosenof}
\date{January 2025}
\usepackage[toc,title,page]{appendix}
\usepackage[margin=1in]{geometry}
\usepackage{multirow}
\usepackage{subcaption}

\usepackage{amsmath}
\DeclareMathOperator\Var{Var}
\DeclareMathOperator\Cov{Cov}

\DeclareMathOperator\F{F}
\DeclareMathOperator\G{G}
\DeclareMathOperator\HFunc{H}
\DeclareMathOperator\E{E}
\DeclareMathOperator\MEV{MEV}
\DeclareMathOperator\MVAR{MVAR}

\usepackage{url}
\usepackage{booktabs}

\newtheorem{lemma}{Lemma}
\newtheorem{assumption}{Assumption}

\begin{document}

\maketitle

\begin{center}
\includegraphics[scale = 0.4]{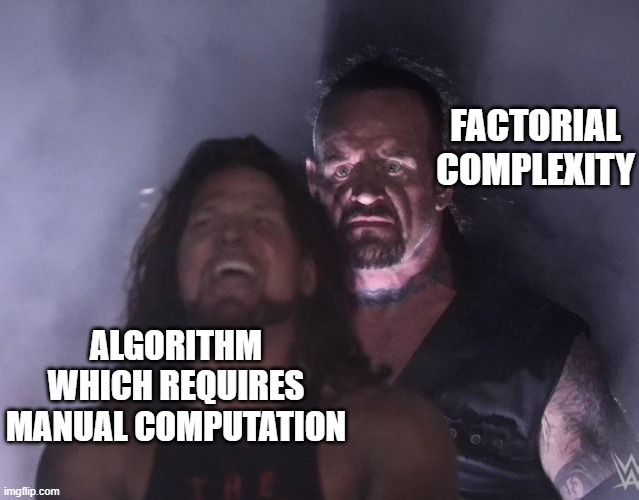}
\end{center}

\abstract{Previous work on fantasy basketball has established methods for optimizing team construction for head-to-head formats. This has been facilitated by the straightforwardness of calculating the objective function for those formats, given that underlying performance distributions are known. Rotisserie has not been optimized in the same way because even with the assumption that performance distributions are known, directly calculating the most natural objective function is intractable. This work introduces a system for making a tractable approximation of that objective function. The resulting simplified objective function aligns well with the traditional wisdom that balanced teams are preferable for the format, because it contains an implicit mechanism that rewards teams for being balanced. Integrating this new objective function into established optimization methods is shown to perform well in the context of simulated seasons.}

\section{Introduction}

Rotisserie leagues have not yet been extensively studied from a mathematical perspective. Two heuristics for quantifying player value have been developed- Z-score and SGP- but they are both fundamentally limited because they do not account for drafting context. A more mathematically sophisticated approach which accounts for drafting context has not been developed.

Recent work introduced the H-scoring framework and the $H_0$ algorithm for selecting optimal players for head-to-head formats (Rosenof, 2024b). The $H_0$ algorithm includes dynamic mechanisms that incorporate drafting context, offering a potential improvement over traditional heuristics if adapted to Rotisserie. However, $H_0$ cannot be directly adapted to Rotisserie because it requires a format-specific objective function. While objective functions have been developed for head-to-head formats, none has been developed for Rotisserie. It is more difficult to develop one for Rotisserie because the probability of winning a Rotisserie league, which would be a natural choice of objective function, is intractable and therefore not feasible to use in practice. 

This work proposes a tractable alternative objective function for the Rotisserie format. Instead of modeling the victory probability directly, it approximates the victory probability under a simplified model of Rotisserie. This allows $H_0$ to be applied to Rotisserie, albeit with some loss of precision

\section{The Rotisserie Format}

The Rotisserie format was invented in 1980 by magazine writer Daniel Okrent for fantasy baseball (Berry, 2020). It is so-called because Okrent's first group of managers often met at ``La Rotisserie Francaise'' in New York City. The format is still popular today and played for other sports in addition to baseball, including basketball (Barutha, 2024).

Like other kinds of fantasy leagues, Rotisserie leagues begin with an auction or draft through which managers select players for their teams. During the fantasy season, which is generally the majority of a professional season, these teams accrue scores across categories based on how their players perform. At the end of the season, teams are ranked for each category and awarded fantasy points accordingly. First place in a category gets $N$ fantasy points, second place gets $N-1$ fantasy points, and so on. The team which earns the most total fantasy points across categories wins the league.

Fantasy points are usually allocated such that the first place team in a $K$-team league earns $K$ fantasy points. In the mathematical sections of this work $K-1$ will be used instead, so that each team is awarded one fantasy point for each team which they surpass in a category

\section{Existing methodologies}

The two traditional methods for player valuation in Rotisserie leagues are Z-scores and Standing Gain Points (SGP) (Ferdinand, 2019). Z-scores, which have been addressed in previous work, are nearly optimal for a highly simplified version of Rotisserie (Rosenof, 2024a). In contrast, SGP takes a fundamentally different approach, relying on empirical observations rather than theoretical models. It uses historical Rotisserie league data to estimate how much of a category is needed to gain one fantasy point in that category (Eassom, 2019). This empirical approach has the advantage of incorporating real-world factors that are difficult to model, such as managers frequently switching players throughout the season. However, its reliance on historical data from comparable leagues can be a limitation, as such data may not always be readily available.

It is known that all static ranking systems are fundamentally flawed for category-based leagues, because they cannot account for different drafting situations (Rosenof, 2024a). Z-scores and SGP may be helpful heuristics in some circumstances, but they are static, and therefore inherently limited.

A more sophisticated approach would adapt to drafting context, including category strengths and remaining position requirements. The $H_0$ approach is dynamic in this way; see previous work for a full description (Rosenof, 2024b). The original work on $H_0$ did not present an objective function for Rotisserie, so it can only be applied to head-to-head formats. But if there was an objective function for Rotisserie, $H_0$ could be extended to work for Rotisserie too

\section{Intractability of objective}

For head to head formats, the natural choice of objective function is a team's expected score per scoring period, since the goal is to do well across many separate scoring periods. This is relatively simple to calculate given that underlying distributions are known (and assuming that categories are independent from each other, for Most Categories). There is no analogous simple objective for Rotisserie. The most natural choice of objective for Rotisserie is the probability of winning the league, since that is what most managers want to do. That is simple to define, but not simple to compute.

Performing the computation requires calculating the probabilities of each individual winning scenario for the team in question, where every possible ordering of teams across all categories is a scenario. Given $|T|$ teams and $|C|$ categories, the number of winning scenarios is approximately 

$$
\frac{(|T|!)^{|C|}}{|T|}
$$

This is because there are $|T|!$ possible orderings for each of $|C|$ categories, of which approximately one in every $|T|$ represents a win for the team.  With 12 teams and nine categories, this value is above $10^{77}$. No modern computer is capable of performing so many operations. This objective function therefore cannot be directly incorporated into $H_0$; a simplification is required

\section{A model and solution for Rotisserie}

It is contended that based on the following simplified model of Rotisserie, the subsequently defined value $V$ is a tractable and differentiable approximation of the probability of a team winning a Rotisserie league. Therefore, it is a sensible objective function for a Rotisserie version of $H_0$

\subsection{Model} \label{Model}

This model extends the assumptions and definitions underlying the logic of $H_0$ with additional assumptions and definitions unique to the Rotisserie format.

\subsubsection{Additional assumptions}

\begin{assumption}  \label{Assumption1} 
Fantasy point totals scored by each team are Normal distributions
\end{assumption}

\begin{assumption}  \label{Assumption2} 
For the purposes of calculating the number of fantasy points needed to win, the distributions of fantasy point totals for opponents are identical and independent
\end{assumption}

\begin{assumption}  \label{Assumption3} 
The distribution of the difference between the maximum number of fantasy points among all opponents and the average number of fantasy points among all opponents is Normal
\end{assumption}

\begin{assumption}  \label{Assumption4} 
When calculating the variance of fantasy points for opposing teams, the expected means of point differentials against opposing teams are distributed independently and Normally. They have a mean of zero, and standard deviation equal to the empirical standard deviation of expected strengths among opposing teams. Also, the distribution of their variance is a Normal distribution
\end{assumption}

The validy of these assumptions is discussed in Section \ref{Limitations}

\subsubsection{Definitions}

\begin{itemize}
\item $\mu_{c,o}$ is the expected X-score mean of team $t$ relative to opponent $o$ in category $c$, divided by $\sqrt{2} \sigma$ where $\sigma$ is the standard deviation of final category totals determined by $H_0$. The normalizing factor is useful because it allows the difference between two teams to have a unit variance
\item $\sigma_c$ is the standard deviation of $\mu_{c,o}$ across opponents for the purposes of the calculation of variance of opposing team fantasy points, multiplied by $\sqrt{2}$. The $\sqrt{2}$ factor is for convenience; it makes it so the distribution of the difference between two opponents has a standard deviation of $\sigma_c$ by the addition of variance
\item $\rho_{a.b}$ is the correlation between a score total for category $a$ and a score total for category $b$
\item $C$ is the set of categories. $|C|$ is the number of categories
\item $O$ is the set of opposing teams. $|O|$ is the number of opponents, or the the number of teams in the league minus one
\item $\Phi$ and $\phi$ are the CDF and PDF of the standard Normal distribution, respectively
\end{itemize}

\subsection{Approximations} \label{Approximations}

To approximate the victory probability based on the model, several mathematical approximations are required. Justifications for these approximations are included in Appendix \ref{apdx.Lemmas}

\begin{lemma}  \label{Approx1} 
So long as $\rho$ is small, the binomial CDF of $(x,y,\rho)$ can be approximated as 

$$
\Phi(x) \Phi(y) + \rho \phi(x) \phi(y)
$$
\end{lemma}

\begin{lemma}  \label{Approx2} 
The maximum of $N$ identical standard Normal distributions has expected value $\MEV(N)$ and variance $\MVAR(N)$ for which values are known. Values up to $N= 20$ are included in Appendix \ref{apdx.Lemmas2}

\end{lemma}

\begin{lemma} \label{Approx 3}

The square root of a large and almost always positive Normal distribution $X$ near its mean is approximately a Normal distribution with mean equal to the square root of the mean of $X$

\end{lemma}

\subsection{Derived objective} \label{Obj}

The resulting Rotisserie objective can be described as a system of equations. For clarity, the equations are separated out into equations representing derived statistical properties of relevant quantities, and helper functions which make those equations simpler to write and compute. A derivation of this system of equations is included in Appendix \ref{apdx.Proof}

\subsubsection{Statistical properties} 

\begin{equation} \label{VProb}
V = \Phi \left(\frac{\mu_D}{\sigma_D} \right)
\end{equation}

\begin{equation} \label{MUD}
\mu_D = \mu_T \frac{|O| + 1}{|O|}- \frac{|C| (|O| + 1) }{2} - \mu_L
\end{equation}

\begin{equation} \label{SUD}
\sigma^2_D = \frac{|O| + 1}{|O|} \sigma^2_T + \sigma^2_L
\end{equation}

\begin{equation} \label{Mean}
\mu_T = \sum_{c \in C} \sum_{o \in O}  \Phi(\mu_{c,o})
\end{equation}

\begin{equation} \label{MUL}
\mu_L = \MEV(|O|) * \sqrt{\E(\sigma^2_M)} 
\end{equation}

\begin{equation} \label{Final Variance}
\sigma^2_T = \sum_{c \in C} \sum_{o \in O} \Phi(\mu_{c,o}) \left( 1 - \Phi(\mu_{c,o}) \right) + \frac{1}{2} \sum_{a \in C} \sum_{b \in C} \rho_{a.b} \HFunc_T(a,b) 
\end{equation}

\begin{equation} \label{SigmaL}
\sigma^2_L =  \E( \sigma^2_M ) * \MVAR(|O|)
\end{equation}

\begin{equation} \label{Generic Variance}
\E(\sigma^2_M) = |O| \sum_{c \in C} \frac{\cos^{-1} \left( \frac{\sigma_c^2}{1 + \sigma_c^2} \right) }{2 \pi}  + \frac{1}{2} \sum_{a \in C} \sum_{b \in C} \rho_{a.b} \HFunc_M(a,b) 
\end{equation}

$V$ is the probability of team $t$ winning the league. $T$ represents team $t$'s own fantasy point total, $M$ represents the distribution of a generic opponent, $D$ represents the difference in total fantasy points between team $t$ and the highest-scoring opponent, and $L$ represents the difference between the highest-scoring opponent and an average opponent. $\mu$ and $\sigma^2$ represent the means and variances of those quantities

\subsubsection{Helper functions}

\begin{equation} \label{DefF}
\F_T(c) = \sum_{o \in O} \phi \left( \mu_{c,o}\right) 
\end{equation}

\begin{equation} \label{DefG}
\G_T(a,b) = \sum_{o \in O} \phi \left( \mu_{a,o}\right) \phi \left( \mu_{b,o}\right) 
\end{equation}

\begin{equation} \label{DefH}
\HFunc_T(a,b) = \left\{
\begin{array}{ll}
      \F_T(c)^2 -  \G_T(c,c) & a = b = c \\[5pt]
      \F_T(a) \F_T(b) +  \G_T(a,b) & a \neq b\\
\end{array} 
\right. 
\end{equation}

\begin{equation} \label{GM}
\HFunc{}_M(a,b) = \frac{|O|}{2  \pi} * \left\{
 \begin{array}{ll}
      \frac{|O| - 1}{\sigma_c^2 + 1 }  & a = b = c\\[10pt]
       \frac{|O| + 1 }{\sqrt{\left( \sigma_a^2 + 1 \right) \left( \sigma_b^2 + 1 \right) } } & a \neq b\\
\end{array} 
\right. 
\end{equation}

\subsubsection{Gradient}

For the purposes of performing H-scoring using gradient descent, it is necessary for the objective function to be differentiable with respect to the underlying values of $\mu_{c,o}$. Indeed, the Rotisserie objective described in Section \ref{Obj} is differentiable. The gradient can be described with the following equations 

\begin{equation} \label{Grad}
\nabla_{c,o} (V) = \frac{\phi \left(\frac{\mu_D}{\sigma_D} \right) }{\sigma_D^3 } \left( \sigma_D^2 * \nabla_{c,o} \left( \mu_D \right) - \frac{\mu_D}{2} *  \nabla_{c,o} \left( \sigma^2_T\right) \right)
\end{equation}

\begin{align}
    \nabla_{c,o} \left( \sigma_T^2 \right) = & \mu_{c,o} \phi(\mu_{c,o}) \left[ \left( \sum_{b \in C \neq c} \rho_{b,c}  \left[ - \phi(\mu_{b,o}) -  F_T(b)\right] \right) +  \left(\phi(\mu_{c,o}) - F_T(c) \right) \right]\nonumber \\ 
    &  + \phi(\mu_{c,o}) - 2 \Phi(\mu_{c,o}) \phi(\mu_{c,o}) \label{SigPGrad}
\end{align}

\begin{equation} \label{mudGrad}
\nabla_{c,o} \left( \mu_D \right) = \frac{|O| + 1}{|O|} \phi(\mu_{c,o}) 
\end{equation} 

A derivation of this gradient is included in Appendix \ref{apdx.Grad}

\section{Simulation}

Simulated versions of NBA fantasy seasons, from 2004-05 to 2023-24, were run to provide reassurance that $H_0$ with objective function defined in Section \ref{Obj} is appropriate for Rotisserie. 

The seasons were simulated in the same way as previous work on H-scoring, except that Rotisserie scoring was used on full-season weekly averages (equivalent to full-season totals), and Gaussian noise was added to categorical performances (Rosenof, 2024b).The covariance of the noise was constructed in the following way

\begin{enumerate}
\item Standard deviations per category were calculated as $\tau_M * |N|$ for counting statistics and $\frac{\tau_R}{|N|}$ for percentage statistics, where $|N|$ is the number of players per team. This corresponds with week-to-week variance of team-level statistics
\item The standard deviations were scaled by $\chi^2$. where $\chi$ was $0.25$, $0.5$, or $0.75$. The value of $\chi$ encodes the confidence in pre-season projections relative to week-to-week variance. For example, $\chi = 0.5$ suggests that if players were scoring plus or minus ten points per week, pre-season forecasts were off by plus or minus five. 
\item $\rho$ was calculated as the average correlation matrix for players in $Q$, with percentage statistics volume-adjusted. Covariance between categories $a$ and $b$ was then calculated as $\rho_{a,b}$ multiplied by both of their respective standard deviations
\end{enumerate}

Both H-scores and G-scores were calculated using the appropriate assumption that full-season variance was equal to week-to-week variance scaled by $\chi$.  

\begin{figure}[ht]

\begin{subfigure}{1\linewidth}
\centering 
\includegraphics[scale = 0.33]{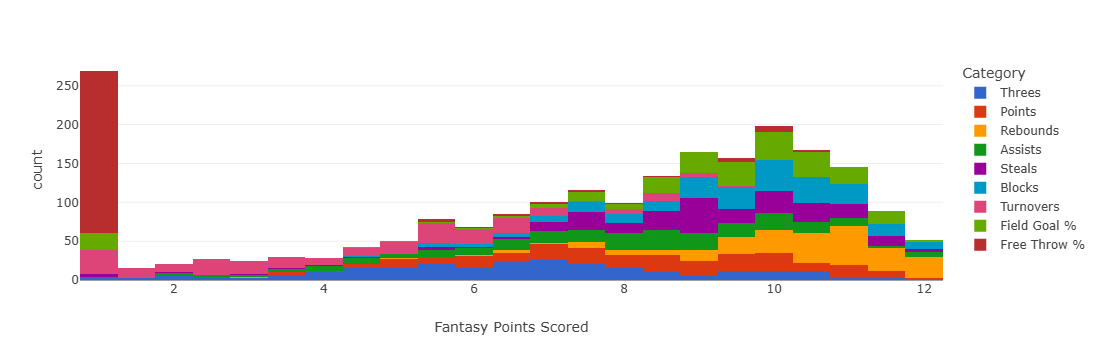}
\caption{$\chi = 0.25$}
\label{RotoCats025}
\end{subfigure}

\begin{subfigure}{1\linewidth}
\centering 
\includegraphics[scale = 0.33]{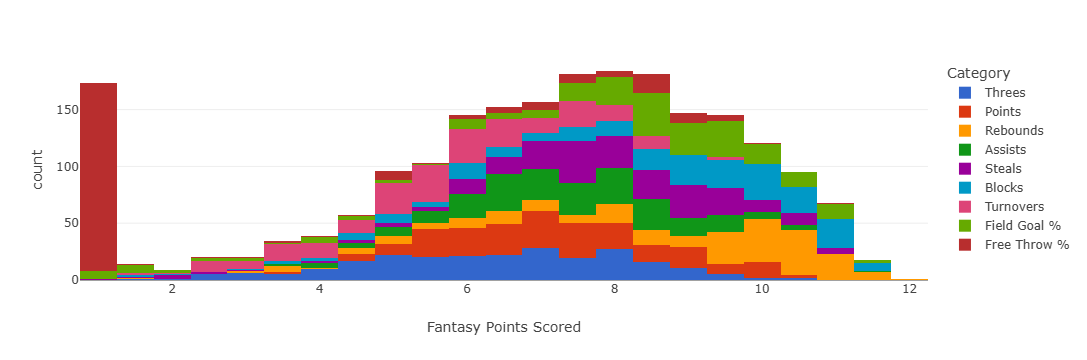}
\caption{$\chi = 0.5$}
\label{RotoCats05}
\end{subfigure}

\begin{subfigure}{1\linewidth}
\centering 
\includegraphics[scale = 0.33]{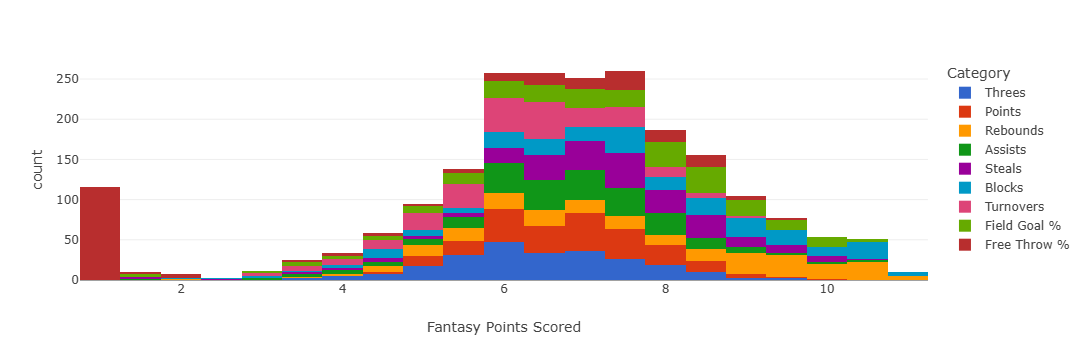}
\caption{$\chi = 0.75$}
\label{RotoCats075}
\end{subfigure}

\caption{H-score drafters' average fantasy points per category. Standard fantasy point scoring is  used for the x-axis, where last place in a category earns one point}
\label{RotoCats}
\end{figure}

The resulting win rates are shown in Table \ref{HvZ}. Figure \ref{RotoCats} shows corresponding fantasy points by category

\begin{table}[ht]
\centering 

\scalebox{0.7}{
\begin{tabular}{llrrrrrrrrrrrrr}
\toprule
 & 0 & 1 & 2 & 3 & 4 & 5 & 6 & 7 & 8 & 9 & 10 & 11 & Mean \\
\midrule
\multirow[t]{21}{*}{$\chi = 0.25$} & 2004-05 & 73.8\% & 48.8\% & 50.4\% & 43.6\% & 67.3\% & 45.9\% & 39.4\% & 38.4\% & 62.0\% & 35.0\% & 64.8\% & 62.3\% & 52.7\% \\
 & 2005-06 & 57.0\% & 33.0\% & 32.1\% & 33.7\% & 61.2\% & 58.3\% & 13.4\% & 12.7\% & 20.9\% & 31.9\% & 27.6\% & 20.8\% & 33.5\% \\
 & 2006-07 & 31.8\% & 20.4\% & 25.2\% & 40.9\% & 45.6\% & 26.1\% & 45.6\% & 38.0\% & 45.8\% & 76.7\% & 61.7\% & 51.1\% & 42.4\% \\
 & 2007-08 & 66.5\% & 50.7\% & 48.9\% & 42.3\% & 40.6\% & 40.5\% & 17.3\% & 18.2\% & 20.6\% & 31.2\% & 15.8\% & 13.7\% & 33.8\% \\
 & 2008-09 & 50.3\% & 65.2\% & 65.2\% & 32.2\% & 63.3\% & 48.3\% & 26.2\% & 25.1\% & 39.2\% & 37.1\% & 34.9\% & 34.9\% & 43.5\% \\
 & 2009-10 & 85.2\% & 73.2\% & 27.5\% & 46.4\% & 43.7\% & 36.6\% & 41.1\% & 41.0\% & 49.3\% & 19.4\% & 27.8\% & 39.4\% & 44.2\% \\
 & 2010-11 & 46.0\% & 51.0\% & 60.7\% & 45.0\% & 34.3\% & 23.4\% & 31.9\% & 51.8\% & 35.3\% & 33.2\% & 38.4\% & 42.4\% & 41.1\% \\
 & 2011-12 & 77.2\% & 34.1\% & 17.8\% & 30.1\% & 21.8\% & 23.6\% & 21.3\% & 20.6\% & 21.6\% & 22.9\% & 20.6\% & 20.7\% & 27.7\% \\
 & 2012-13 & 60.1\% & 36.2\% & 21.0\% & 18.4\% & 20.9\% & 11.8\% & 23.4\% & 29.0\% & 24.0\% & 17.2\% & 18.6\% & 20.1\% & 25.1\% \\
 & 2013-14 & 53.7\% & 69.6\% & 48.3\% & 40.4\% & 32.2\% & 35.6\% & 28.5\% & 12.2\% & 17.9\% & 21.5\% & 18.6\% & 24.1\% & 33.5\% \\
 & 2014-15 & 59.2\% & 71.6\% & 66.0\% & 23.6\% & 24.8\% & 16.3\% & 12.4\% & 16.0\% & 19.4\% & 42.6\% & 41.6\% & 41.8\% & 36.3\% \\
 & 2015-16 & 72.1\% & 27.9\% & 35.2\% & 17.5\% & 20.7\% & 27.9\% & 25.9\% & 27.7\% & 24.4\% & 34.8\% & 37.0\% & 38.3\% & 32.5\% \\
 & 2016-17 & 24.0\% & 19.4\% & 14.2\% & 46.9\% & 35.3\% & 38.9\% & 35.8\% & 33.1\% & 28.0\% & 30.8\% & 39.5\% & 56.6\% & 33.5\% \\
 & 2017-18 & 65.6\% & 48.9\% & 54.4\% & 48.8\% & 32.5\% & 17.1\% & 18.4\% & 18.5\% & 18.5\% & 27.4\% & 31.6\% & 22.5\% & 33.7\% \\
 & 2018-19 & 47.7\% & 48.4\% & 42.1\% & 49.6\% & 20.6\% & 24.2\% & 21.1\% & 40.8\% & 34.1\% & 32.9\% & 25.1\% & 23.8\% & 34.2\% \\
 & 2019-20 & 39.4\% & 33.6\% & 36.2\% & 48.5\% & 43.8\% & 46.7\% & 40.0\% & 46.1\% & 47.7\% & 31.2\% & 37.0\% & 36.2\% & 40.5\% \\
 & 2020-21 & 39.6\% & 32.7\% & 33.7\% & 34.5\% & 32.6\% & 62.9\% & 65.4\% & 84.3\% & 83.4\% & 81.8\% & 61.5\% & 78.3\% & 57.6\% \\
 & 2021-22 & 69.3\% & 45.3\% & 49.1\% & 21.7\% & 38.2\% & 37.0\% & 31.7\% & 38.8\% & 34.2\% & 34.6\% & 64.4\% & 43.0\% & 42.3\% \\
 & 2022-23 & 34.4\% & 46.3\% & 26.9\% & 36.8\% & 39.3\% & 42.8\% & 23.8\% & 33.2\% & 34.8\% & 25.8\% & 46.2\% & 26.4\% & 34.7\% \\
 & 2023-24 & 33.7\% & 30.4\% & 25.8\% & 27.2\% & 28.4\% & 27.5\% & 46.0\% & 18.3\% & 15.0\% & 25.1\% & 26.4\% & 28.2\% & 27.7\% \\
 & Mean & 54.3\% & 44.3\% & 39.0\% & 36.4\% & 37.4\% & 34.6\% & 30.4\% & 32.2\% & 33.8\% & 34.7\% & 37.0\% & 36.2\% & 37.5\% \\
\midrule
\multirow[t]{21}{*}{$\chi = 0.5$} & 2004-05 & 24.0\% & 23.3\% & 23.1\% & 28.4\% & 19.5\% & 25.3\% & 21.4\% & 18.6\% & 23.2\% & 22.0\% & 16.9\% & 16.2\% & 21.8\% \\
 & 2005-06 & 19.1\% & 20.6\% & 20.0\% & 21.7\% & 19.6\% & 21.9\% & 16.4\% & 10.7\% & 11.3\% & 7.5\% & 7.5\% & 11.3\% & 15.6\% \\
 & 2006-07 & 18.4\% & 17.9\% & 17.5\% & 20.9\% & 20.0\% & 21.2\% & 20.9\% & 16.4\% & 18.9\% & 17.2\% & 23.6\% & 23.8\% & 19.7\% \\
 & 2007-08 & 17.9\% & 14.7\% & 26.6\% & 15.7\% & 17.7\% & 22.0\% & 18.6\% & 23.0\% & 19.8\% & 12.6\% & 18.1\% & 13.7\% & 18.4\% \\
 & 2008-09 & 28.6\% & 25.3\% & 28.4\% & 17.3\% & 13.0\% & 7.1\% & 16.2\% & 19.3\% & 16.8\% & 15.7\% & 16.2\% & 20.5\% & 18.7\% \\
 & 2009-10 & 32.3\% & 23.6\% & 14.6\% & 20.4\% & 23.8\% & 24.3\% & 21.8\% & 20.9\% & 16.1\% & 13.0\% & 12.9\% & 19.4\% & 20.2\% \\
 & 2010-11 & 23.4\% & 22.9\% & 23.5\% & 17.2\% & 19.3\% & 19.3\% & 12.3\% & 13.7\% & 11.3\% & 20.4\% & 18.9\% & 17.8\% & 18.3\% \\
 & 2011-12 & 23.1\% & 41.2\% & 23.2\% & 16.4\% & 16.8\% & 12.4\% & 15.0\% & 14.6\% & 16.8\% & 15.4\% & 16.0\% & 14.6\% & 18.8\% \\
 & 2012-13 & 21.7\% & 26.7\% & 20.5\% & 8.5\% & 9.3\% & 13.1\% & 8.0\% & 7.7\% & 6.5\% & 6.6\% & 9.8\% & 7.7\% & 12.2\% \\
 & 2013-14 & 24.1\% & 10.2\% & 7.6\% & 12.2\% & 8.0\% & 10.5\% & 15.6\% & 8.4\% & 13.4\% & 13.3\% & 17.2\% & 9.8\% & 12.5\% \\
 & 2014-15 & 27.9\% & 25.2\% & 22.4\% & 25.5\% & 11.1\% & 13.0\% & 14.5\% & 15.8\% & 16.3\% & 19.4\% & 21.6\% & 20.2\% & 19.4\% \\
 & 2015-16 & 38.3\% & 24.0\% & 17.9\% & 18.7\% & 20.3\% & 19.2\% & 17.4\% & 19.7\% & 21.4\% & 19.0\% & 20.0\% & 23.8\% & 21.6\% \\
 & 2016-17 & 13.5\% & 12.5\% & 11.4\% & 13.5\% & 13.4\% & 13.1\% & 11.7\% & 6.4\% & 6.4\% & 6.2\% & 6.5\% & 18.3\% & 11.1\% \\
 & 2017-18 & 23.6\% & 16.9\% & 19.1\% & 14.0\% & 16.5\% & 6.9\% & 13.7\% & 8.1\% & 8.5\% & 8.5\% & 10.7\% & 8.1\% & 12.9\% \\
 & 2018-19 & 20.2\% & 21.1\% & 18.9\% & 19.7\% & 14.6\% & 14.0\% & 14.1\% & 19.4\% & 17.8\% & 15.2\% & 19.6\% & 14.4\% & 17.4\% \\
 & 2019-20 & 23.1\% & 13.4\% & 18.2\% & 18.9\% & 18.1\% & 18.9\% & 21.4\% & 17.3\% & 18.7\% & 18.3\% & 18.8\% & 18.3\% & 18.6\% \\
 & 2020-21 & 29.9\% & 15.6\% & 18.7\% & 13.9\% & 16.5\% & 20.0\% & 20.1\% & 19.0\% & 19.9\% & 19.4\% & 23.6\% & 17.3\% & 19.5\% \\
 & 2021-22 & 22.4\% & 17.0\% & 15.4\% & 20.9\% & 20.2\% & 21.5\% & 17.9\% & 19.2\% & 20.9\% & 22.7\% & 17.1\% & 19.0\% & 19.5\% \\
 & 2022-23 & 14.1\% & 9.3\% & 24.2\% & 11.2\% & 8.8\% & 12.0\% & 9.8\% & 9.7\% & 11.3\% & 14.0\% & 17.0\% & 18.2\% & 13.3\% \\
 & 2023-24 & 14.1\% & 14.0\% & 14.6\% & 8.0\% & 19.8\% & 9.4\% & 13.2\% & 14.1\% & 14.6\% & 14.8\% & 13.1\% & 15.8\% & 13.8\% \\
 & Mean & 23.0\% & 19.8\% & 19.3\% & 17.1\% & 16.3\% & 16.3\% & 16.0\% & 15.1\% & 15.5\% & 15.1\% & 16.2\% & 16.4\% & 17.2\% \\
\midrule
 \multirow[t]{21}{*}{$\chi = 0.75$} & 2004-05 & 12.0\% & 13.1\% & 12.2\% & 13.1\% & 12.4\% & 12.6\% & 12.3\% & 12.5\% & 14.9\% & 10.6\% & 11.2\% & 10.0\% & 12.2\% \\
 & 2005-06 & 13.6\% & 13.7\% & 16.3\% & 15.1\% & 12.7\% & 11.2\% & 8.4\% & 9.5\% & 9.2\% & 7.8\% & 9.8\% & 9.4\% & 11.4\% \\
 & 2006-07 & 13.5\% & 12.9\% & 12.0\% & 12.1\% & 12.3\% & 12.6\% & 11.9\% & 10.7\% & 11.2\% & 11.2\% & 15.2\% & 14.8\% & 12.5\% \\
 & 2007-08 & 15.8\% & 11.0\% & 14.0\% & 17.3\% & 15.6\% & 13.1\% & 12.6\% & 17.4\% & 7.8\% & 9.4\% & 9.5\% & 8.3\% & 12.7\% \\
 & 2008-09 & 14.5\% & 22.9\% & 20.3\% & 9.3\% & 12.0\% & 12.7\% & 11.8\% & 13.1\% & 9.8\% & 10.2\% & 9.6\% & 10.1\% & 13.0\% \\
 & 2009-10 & 15.9\% & 15.7\% & 12.7\% & 15.5\% & 15.3\% & 16.0\% & 13.9\% & 15.7\% & 10.6\% & 9.6\% & 10.1\% & 9.9\% & 13.4\% \\
 & 2010-11 & 15.6\% & 13.1\% & 13.1\% & 11.6\% & 12.2\% & 13.2\% & 16.4\% & 13.7\% & 13.7\% & 10.8\% & 12.3\% & 13.1\% & 13.2\% \\
 & 2011-12 & 20.5\% & 21.9\% & 14.9\% & 11.5\% & 10.4\% & 8.6\% & 14.3\% & 10.5\% & 10.8\% & 11.3\% & 11.8\% & 9.9\% & 13.1\% \\
 & 2012-13 & 16.5\% & 16.7\% & 8.2\% & 9.3\% & 7.8\% & 11.8\% & 10.7\% & 8.3\% & 8.7\% & 12.7\% & 13.1\% & 8.1\% & 11.0\% \\
 & 2013-14 & 15.2\% & 11.6\% & 10.0\% & 11.5\% & 8.7\% & 8.6\% & 8.2\% & 8.9\% & 8.5\% & 10.9\% & 10.3\% & 9.1\% & 10.1\% \\
 & 2014-15 & 19.0\% & 13.9\% & 13.6\% & 15.5\% & 14.1\% & 13.0\% & 12.8\% & 12.9\% & 9.3\% & 7.9\% & 12.0\% & 15.4\% & 13.3\% \\
 & 2015-16 & 26.3\% & 17.8\% & 10.7\% & 16.2\% & 13.5\% & 13.9\% & 15.7\% & 16.5\% & 14.7\% & 14.5\% & 16.9\% & 14.6\% & 15.9\% \\
 & 2016-17 & 10.9\% & 12.6\% & 11.2\% & 13.8\% & 12.6\% & 9.6\% & 10.1\% & 8.2\% & 7.1\% & 6.0\% & 4.9\% & 9.4\% & 9.7\% \\
 & 2017-18 & 16.2\% & 12.2\% & 11.3\% & 12.0\% & 7.9\% & 8.5\% & 10.2\% & 9.7\% & 8.6\% & 8.7\% & 8.1\% & 5.1\% & 9.9\% \\
 & 2018-19 & 16.1\% & 12.4\% & 11.1\% & 16.5\% & 9.1\% & 11.4\% & 12.6\% & 11.6\% & 13.5\% & 16.0\% & 13.0\% & 9.0\% & 12.7\% \\
 & 2019-20 & 13.6\% & 11.6\% & 11.9\% & 9.7\% & 9.1\% & 13.5\% & 11.1\% & 13.4\% & 11.5\% & 12.1\% & 11.7\% & 8.5\% & 11.5\% \\
 & 2020-21 & 19.9\% & 17.4\% & 13.0\% & 10.2\% & 15.0\% & 14.4\% & 18.0\% & 12.9\% & 13.0\% & 11.3\% & 13.5\% & 12.7\% & 14.3\% \\
 & 2021-22 & 16.7\% & 12.7\% & 11.1\% & 11.2\% & 9.9\% & 9.4\% & 12.5\% & 12.2\% & 11.7\% & 12.8\% & 10.1\% & 11.8\% & 11.8\% \\
 & 2022-23 & 12.1\% & 9.8\% & 10.2\% & 11.8\% & 8.7\% & 11.6\% & 8.4\% & 10.9\% & 10.0\% & 11.0\% & 9.6\% & 10.0\% & 10.4\% \\
 & 2023-24 & 13.8\% & 12.0\% & 12.5\% & 8.4\% & 11.2\% & 6.7\% & 9.9\% & 8.6\% & 9.3\% & 10.8\% & 7.2\% & 11.9\% & 10.2\% \\
 & Mean & 15.9\% & 14.3\% & 12.5\% & 12.6\% & 11.5\% & 11.6\% & 12.1\% & 11.9\% & 10.7\% & 10.8\% & 11.0\% & 10.6\% & 12.1\% \\
\bottomrule
\end{tabular}

}

\caption{Rotisserie Win rates for $H_0$ against a field of G-score drafters, by year and draft seat}
\label{HvZ}
\end{table}

\section{Discussion}

\subsection{Simulation methodology}

Previous work has simulated head-to-head formats by sampling weekly results from a real season, with the synthetic managers having full knowledge of the underlying distributions (Rosenof, 2024a). This set-up is reasonable for head-to-head formats because the sampling process replicates week-to-week variance. Theoretically, real head-to-head match-ups have additional variance because of projection inaccuracy, but not so much as to make the simulations wholly unrepresentative. 

The same cannot be said of Rotisserie. The only source of variance for Rotisserie is inaccuracy of pre-season projections. Excluding that source of variance would make the results deterministic, which would not represent real Rotisserie well.

Two bounds on the variance of pre-season projections are clear intuitively- it is positive, and likely below week-to-week variance. Unfortunately, as far as this author is aware, there has been no comprehensive survey of the accuracy of pre-season forecasts. This motivates the use of the $\chi$ parameter to encode forecast accuracy as somewhere on the spectrum between zero and week-to-week variance.

One might also question the validity of using player-level averages for $\rho$. The reasoning behind it is that covariance and variance both scale linearly under addition of independent variables. Therefore correlation, which is a ratio between covariance and variance, does not scale at all. This means that if players truly were chosen randomly, the expected correlation of their statistical sums would be equal to their individual expected correlations

\subsection{Simulation results}

Table \ref{HvZ} shows that $H_0$ with the Rotisserie objective performed well against a field of G-score agents, especially when $\chi$ was low and the effect of random chance was minimal. It won $37.5\%$, $17.2\%$, and $12.1\%$ of its seasons with $\chi$ set to $0.25$, $0.5$, and $0.75$ respectively. These are all above the baseline expected rate of $\frac{1}{12} = 8.3\%$

Figure \ref{RotoCats} shows that the Rotisserie version of H-scoring did not punt (strategically abandon a category) much, particularly when compared to head-to-head versions (Rosenof, 2024b). However, it did punt Free Throw \% on occasion, especially when $\chi$ was low. 

This behavior tracks well with the traditional strategy for Rotisserie, which is to punt minimally. As NBC Sports summarizes, ``Punting is best left to weekly head-to-head leagues ... [Rotisserie] league managers should only consider the approach for one category. And that would be under extreme circumstances (NBC Sports, 2024).'' 

\subsubsection{Why punt less often?} \label{Intuition}

There is an established rationale, based on intuition, for why punting is a poor choice for Rotisserie leagues (Lamdin, 2015). The idea is that punting decreases the margin of error for other categories. Say that a typical winning Rotisserie team averages third place out of twelve across nine categories. A single punted category sacrifices eleven fantasy points out of the eighteen that a team can afford to lose, forcing them to earn an average placement between first and second in every other category in order to win. This is a difficult level of dominance to achieve, even with the boost from punting. 

This argument can also be framed mathematically. In Rotisserie, a team needs an exceptional ``upside'' performance to surpass all other teams and win. The probability of this happening is greatly influenced by the variance of a team's fantasy point total. With low variance, they are unlikely to get any kind of extreme performance, including the kind of upside performance required to win. With high variance, they are relatively more likely to score on the extreme at either end, making an upside performance more realistic. So teams are more likely to win overall with higher variance in their fantasy point totals. Punting reduces the variance of fantasy point totals because it increases clarity on which fantasy points will be won and which will be lost, narrowing the spread of outcomes. Therefore, one would expect punting to decrease the probability of overall victory.

This mathematical reasoning can be found within the formulas of Section \ref{Obj}. $\mu_D$ is generally negative, so as $\sigma_D$ increases, $\frac{\mu_D}{\sigma_D}$ becomes a smaller negative number and $V$ increases. $\sigma_D$ depends on the variance of the team's own point distribution. The variance of a potential ranking point as encapsulated within Equation \ref{Final Variance} is $\Phi (1- \Phi)$, where $\Phi$ is the probability of winning the underlying matchup. It is maximized when $\Phi=\frac{1}{2}$. Therefore, all else held equal, the manager would prefer its match-ups to be as close to 50-50 as possible. 

Of course, the incentive to punt still exists from the perspective of maximizing the expected value of match-up wins. It is possible for punting to increase overall expected value so much that it counteracts the negative implications of decreasing $\sigma_D$

\subsubsection{Why punt Free Throw \%?} 

The Free Throw \% category is particularly conducive to punting because there is a group of players who are generally strong but hindered by exceptionally poor free throw shooting. In the $\chi = 0.25$ simulations, every H-score team that punted free throws selected at least one from a set of four notoriously poor free throw shooters, as shown in Table \ref{FTPunts}. In the simulated drafts, the extreme Free Throw deficiencies of these players reduced their total G-scores, encouraging the pure G-score drafters to avoid them and leave them for the H-score drafter. In practice, the algorithm's tendency to punt free throws will depend on the availability of these unique players

\begin{table}
\centering
\begin{tabular}{| l | l |} 
\hline
Players & Team count \\
\hline
Dwight Howard & 56 \\
Giannis Antetokounmpo & 50 \\
Dwight Howard / Shaquille O'Neal & 33 \\
Andre Drummond / Dwight Howard & 28 \\
Giannis Antetokounmpo / Andre Drummond & 13 \\
Andre Drummond & 13 \\
Shaquille O'Neal & 11 \\
Giannis Antetokounmpo / Dwight Howard & 3 \\
\hline
Total & 207 \\
\hline
\end{tabular}
\caption{Table of players on free-throw punting teams, defined by those with average scores below  1.5 points in the Free Throw \% category. From the $\chi=0.25$ simulations. All of the 207 free throw punting teams had at least one of Dwight Howard (49 to 67\% free throw shooter), Shaquille O'Neal (42-62\%), Giannis Antetokounmpo (61-77\%), or Andre Drummond (35-61\%). The NBA average free throw percent is currently 78\%}
\label{FTPunts}
\end{table}

\subsubsection{Why punt more when when $\chi$ is lower?}

When $\chi$ is high, the probability of winning any fantasy point will not deviate far from $50\%$. This disrupts the logic of punting- it is not worth completely abandoning a category if there is still some chance of winning it, and the advantages gained in other categories would be marginal anyway. 

On the flip side, with high values of $\chi$, the algorithm is both more confident that it will lose the fantasy points of a weak category, and more confident than advantages in other categories will lead to consistent over-performance. This encourages it to punt the weak categories

\subsection{Limitations} \label{Limitations}

The Rotisserie objective described in Section \ref{Obj} is only valid to the extent that the assumptions laid out in Section \ref{Model} are valid. Their imperfections lead to limitations in the resulting algorithm

\subsubsection{Fantasy point totals are distributed Normally}

The most precise way to model the total number of fantasy points for a team would be with a binomial distribution with dependent trials.  This would accurately reflect the discrete nature of fantasy points, but would also have the downside of making analysis difficult, motivating the use of the Normal approximation. 

Approximating the distribution Normally via the central limit theorem is not entirely justified, even with a high number of trials, because the CLT does not apply when trials are correlated. Fortunately, it is known that the CLT can be relaxed to a degree for weak dependence structures (Bradley, 1981). The relaxed CLT does not necessarily apply in this case, but it does offer some hope that introducing correlations does not radically modify the distribution from approximately Normal. 

It is worth noting that the number of fantasy points is usually rather high and correlations are usually rather small. With $|C|$ of eight and $|O|$ of nine (which would represent a smaller than usual league), there would still be 64 possible fantasy points. Most of them would represent matchups against different opponents in different categories, which would be only loosely correlated. So it is perhaps not naive to hope that between the high number of fantasy points and the weakness of many of the correlations between them, applying the CLT is not too problematic

\subsubsection{Opposing teams have identical fantasy point total distributions}

Assumption \ref{Assumption2} dictates that for the purpose of calculating the fantasy point total target, all opposing teams have the same distribution of fantasy points, all independent from each other.

The assumption that the distributions are identical ignores the possibility that some opponents may have teams with above-average strength, making them more likely to be the winner and driving up the expected value of the target. This assumption is likely to be violated often. Even if all opposing drafters are drafting optimally, managers with high draft seats tend to have systematic advantages due to the marginal differences between players being larger at the higher end (Rosenof, 2024b). This means that drafters in high draft seats will likely have stronger than average teams. 

To ameliorate this problem, it would be ideal to incorporate the actual expected values of fantasy point totals into the calculation of the properties of the maximum. Mathematically this would add complexity: unlike $\MEV$ and $\MVAR$. the statistical properties of the maximum of non-identical random variables cannot be computed beforehand. But the properties could perhaps be calculated on the fly, aided by existing literature on the maximum of non-identical distributions (Engelke, 2015). 

The assumption that the fantasy point totals of other teams are independent from each other is also inaccurate, because opponents are competing for the same fantasy points with each other. One opponent doing better means that other opponents must perform worse in aggregate. Ideally this would also be incorporated into the calculation of the maximum, but the maximum of non-independent, non-identical distributions is even more complicated than of purely non-identical distributions: they are not covered in Engelke's work, for example. Tackling this may be significantly non-trivial. 

An alternative to refining the calculation of the maximum is to use empirical results instead. Of course, this has the downside, like SGP, of requiring a historical record of similar leagues. 

Fortunately, all reasonable procedures for estimating the target should lead to objectives with similar properties, even if some are less precise than others. The target should be significantly above the expected fantasy point total, incentivizing the Rotisserie drafter to optimize for upside

\subsubsection{The difference between the average and maximum opponent fantasy point total is distributed Normally}

The maximum of many Normal distributions is approximately a Gumbel distribution, not a Normal distribution. Fortunately, Gumbel distributions are similar to Normal distributions. As one paper investigating them in the context of flood engineering notes, ``the normal and Gumbel distributions are much alike in practical engineering'' (Abdelaziz, 2016).

Using a Normal distribution instead of a Gumbel makes the calculation of the final objective much simpler. Since Gumbels are similar to Normal distributions for practical purposes, this likely does not skew the result much. 

It is also possible that the number of other teams may be insufficient to justify the use of any large-N approximation. This may be a problem for extremely small leagues 

\subsubsection{The variance of the fantasy points totals of opponents can be calculated in a particular way}

Since opposing teams are assumed to have identical distributions by Assumption \ref{Assumption2}, there is a need to estimate their shared variance. Assumption \ref{Assumption4} provides a reasonable way to make that estimation. Essentially it is reducing the space of other teams' categorical strengths to Normal distributions, which is of course not entirely accurate, but is helpful for calculation purposes.

The most potentially objectionable specification of Assumption \ref{Assumption4} is that the manager in question's team is not considered. This is for convenience. It allows the value of $L$ to be calculated independently of the manager's own team and decisions, precluding the need to re-calculate it for each candidate player and on each step of gradient descent. It is also intuitively reasonable; one would not expect the variance of opponents to be greatly influenced by the manager's own choices

\subsubsection{Original assumptions of $H_0$, including that all players count}

One of the core assumptions of the $H_0$ algorithm is that the performances of all drafted players count for the team that drafted them (Rosenof, 2024b). This assumption can be problematic for several reasons, one of which is especially problematic for Rotisserie. Managers do not always consistently set their line-ups, especially when they are not performing well enough to compete for a top placement. At the end of a Rotisserie season, there may be a number of managers who are so far behind that they have effectively no chance to win. They are less likely to set their line-ups properly, thereby falling even further behind on counting statistics. However, these managers would have no disadvantage in the percentage statistics. They could still win fantasy points in them over managers who are actively competing. This perhaps suggests that a manager hoping to perform well across the board should prioritize the percentage statistics, since those will be more difficult to win fantasy points for. One could also make the argument that it makes the counting statistics less attractive to punt, since punting a counting statistic would forfeit the almost-free points to be earned against inattentive managers in that category.

Additionally, all of the other potential issues of H-scoring for head-to-head formats apply to Rotisserie as well. Position requirements are not totally flexible, teams may change because of injuries, etc. These potential issues are significant, and motivate careful human consideration when using the algorithm

\subsubsection{The goal is to win the league}

The objective function is designed as a proxy of the probability of overall victory. Most managers want to win their leagues, but this might not be their only consideration. There may be prizes for other top placements, or punishments for particularly poor performances. An ideal objective function would be able to account for this flexibly

\section{Future work}

There are many areas for future work, including 

\begin{itemize}
\item Modeling Rotisserie with more precision, perhaps by describing the difference between the fantasy point target and the expectation of the average opponent with more precision
\item Improving estimates of the performance uncertainty of pre-season projections. This would allow for better-calibrated X-scores and more realistic simulations of Rotisserie 
\item Customizing the objective function to account for managers who may be interested in second- or third-place finishes. E.g. the reward structure could be 70\% for first place, 20\% for second, and 10\% for third 
\end{itemize}

\section{Conclusion}

A reasonable heuristic that can be used as a Rotisserie objective is presented. It is not as precise as manual computation of the objective, but is much more tractable. It works reasonably well in simulations.

\textit{Disclaimer: The views and opinions expressed in this article are those of the independent author and do not represent those of any organization, company or entity}

\begin{appendices}

\section{Justifying the equations} \label{apdx.Proof}

The system of equations described by Section \ref{Obj} can be justified by analyzing the statistical properties of relevant quantities under the assumptions of Section \ref{Model}. This justification also makes use of the approximations described in Section \ref{Approximations}

\subsection{Team $t$'s fantasy point total}

By Assumption \ref{Assumption1}, team $t$'s fantasy point total is a Normal distribution. Therefore it can be fully parameterized by its mean and variance, represented by $\mu_T$ and $\sigma^2_T$

\subsubsection{Mean}

The expected value of team $t$'s fantasy point total $\mu_T$ is the sum of the probabilities of team $t$ winning each match-up. Since score totals are always Normal distributions by the original assumption of $H_0$ and $\mu_{c,o}$'s are in a basis such that point differentials have a unit variance, team $t$'s probability of winning category $c$ against opponent $o$ is $\Phi (\mu_{c,o})$. Their expected fantasy point total is then the sum of $\Phi (\mu_{c,o})$ across categories and opponents, as represented by Equation \ref{Mean}

\subsubsection{Variance}

The overall variance of team $t$'s fantasy point total is the sum of variances of each potential ranking point, plus two times all of their pairwise covariances.

The variance terms are relatively simple to calculate. The variance of a Bernoulli event is $p \left( 1 - p \right)$. So the variance of a ranking point for a specific category matchup against a specific opponent is 

\begin{equation} \label{VarianceTerm}
\Phi(\mu_{c,o}) \left( 1 - \Phi(\mu_{c,o}) \right)
\end{equation}

Their sum is then

\begin{equation} \label{FullVarianceTerm}
\sum_{c \in C} \sum_{o \in O} \Phi(\mu_{c,o}) \left( 1 - \Phi(\mu_{c,o}) \right)
\end{equation}

Covariance terms must be calculated for every pair of potential fantasy points, of which there are three kinds: same category different opponent, same opponent different category, and different opponent different category. 

Each of these cases can be handled by a general framing. Consider team $t$ competing against teams $m$ and $n$ in categories $a$ and $b$. $a$ and $b$ could be the same and $m$ and $n$ could be the same, but $m$ and $n$ must be different from $t$. The scores of the two match-ups can be represented by a multivariate normal distribution in four dimensions; $A_t$, $A_m$, $B_t$, and $B_n$.

Dub the differential of the first matchup, $A_t - A_m$, as $A$. Dub the differential of the second matchup $B_t - B_n$, as $B$. Say that $A_t$ and $B_t$ have correlation $\rho_1$, and $A_m$ and $B_n$ have correlation $\rho_2$. $A_t$ / $B_n$ and $A_m$ / $B_t$ always have zero correlation because they are associated with different teams. 

By how $\mu_{c,o}$ was defined, $A$ has mean $\mu_{a,m}$ and variance one, and $B$ has mean $\mu_{b.n}$ and variance one. Also, $A_t$, $A_m$ etc. must have variance $\frac{1}{2}$, to be consistent with $A$ and $B$ having a variance of one.

The covariance of two variables with the same variance is equal to their correlation times their shared variance. Therefore, $A_t$ and $B_t$ have a covariance of $\frac{\rho_1}{2}$, and $A_m$ and $B_n$ have a covariance of $\frac{\rho_2}{2}$. 

Covariance is additive, so the covariance between $A$ and $B$ is 

\begin{align*}
& \Cov(A_t, B_t) + \Cov(A_m, B_n) + \Cov(A_t, B_n) + \Cov(A_m, B_t) \\
& = \frac{\rho_1}{2} +  \frac{\rho_2}{2} + 0 + 0 \\
& = \frac{\rho_1 + \rho_2}{2}
\end{align*}

$A$ and $B$ have unit variance, so this is also the correlation between them. 

Team $t$ wins fantasy points based on whether $A$ and $B$ are positive (given that they are Normal distributions, the probability of a tie is infinitesimal). Call $A'$ the point given for $A> 0$ and $B'$ the point given for $B > 0$. 

The covariance of $A'$ and $B'$ is 

$$
Cov(A',B') = E(A' B') - E(A')E(B')
$$

The $E(A')$ and $E(B')$ terms are easy to calculate. They are $\Phi(\mu_A)$ and $\Phi(\mu_B)$. So the equation can be rewritten to 

\begin{equation} \label{FirstCovEq}
Cov(A',B') = E(AB) - \Phi \left( \mu_{a,m} \right) * \Phi \left( \mu_{b,n} \right)
\end{equation}

The probability of both $A'$ and $B'$ occurring simultaneously is the standard bivariate normal CDF at $\mu_A$ and $\mu_B$ . That is, 

$$
BvN \left[\mu_{a,m}, \mu_{b,n}; \rho = \frac{\rho_1 + \rho_2}{2}  \right]
$$

By Lemma \ref{Approx1}, for small values of $\rho$, $\Phi(x) \Phi(y) + \rho \phi(x) \phi(y)$ is a good approximation of the CDF of a standard multivariate Normal at $x$ and $y$. So 

$$
E(AB) \approx \Phi \left( \mu_{a.m} \right) * \Phi \left( \mu_{b,n} \right) + \frac{\rho_1 + \rho_2}{2} \left( \phi \left( \mu_{a.m} \right) * \phi \left( \mu_{b,n} \right) \right) 
$$

Subbing back into Equation \ref{FirstCovEq} yields 

\begin{align*}
Cov(A',B') = 
& \Phi \left( \mu_{a,m} \right) * \Phi \left( \mu_{b,n} \right) + \frac{\rho_1 + \rho_2}{2} \left( \phi \left( \mu_{a,m} \right) * \phi \left( \mu_{b,n} \right) \right) \\
& - \Phi \left( \mu_{a,m} \right) * \Phi \left( \mu_{b,n} \right) \\
= & \frac{\rho_1 + \rho_2}{2} \left( \phi \left( \mu_{a,m} \right) * \phi \left( \mu_{b,n} \right) \right)
\end{align*}

The values of $\rho_1$ and $\rho_2$ are different for each of the three score pair cases

\begin{itemize}
\item Same category different opponent:  $A_t$ and $B_t$ represent the same quantity, so $\rho_1 = 1$. $A_m$ and $B_n$ are from different opponents, so  $\rho_2 = 0$. Thus, 

$$
\frac{\rho_1 + \rho_2}{2} = \frac{1 + 0}{2} = \frac{1}{2}
$$

\item Different category same opponent: $A_t$ and $B_t$ represent the same team across different categories, so $\rho_1$ is $\rho_{A,B}$. $A_m$ and $B_n$ are also the same team across different categories, so $\rho_2$ is also $\rho_{A,B}$.

$$
\frac{\rho_1 + \rho_2}{2} = \frac{\rho_{a,b} + \rho_{a,b}}{2} = \rho_{a,b}
$$

\item Different category different opponent: $\rho_1$ is still $\rho_{a,b}$. However, $A_m$ and $B_n$ are from different opponents, so $\rho_2 = 0$

$$
\frac{\rho_1 + \rho_2}{2} = \frac{\rho_{a,b} + 0}{2} = \frac{\rho_{a,b}}{2}
$$
\end{itemize}

Now all of the covariance terms can be summed. Using the fact that

$$
2 \Cov(X,Y) = \Cov(X,Y) + \Cov(Y, X)
$$

The covariance terms can be summed across all potential pairings, double-counting all of them with no correction factor. Writing them all out explicitly based on the three cases yields:

\begin{align*}
& \sum_{c \in C} \sum_{m \in O}  \sum_{n \in O \neq m} \frac{1}{2} \phi \left( \mu_{c,n} \right) \phi \left( \mu_{c,m} \right)  \\
+ & \sum_{a \in C} \sum_{b \in C \neq a} \sum_{o \in O} \rho_{a.b} * \phi \left( \mu_{a,o} \right) \phi \left( \mu_{b,o} \right)  \\
+ & \sum_{a \in C} \sum_{b \in C \neq a} \sum_{m \in O} \sum_{n \in O \neq m} \frac{\rho_{a.b}}{2} \phi \left( \mu_{a,m} \right) \phi \left( \mu_{b,n} \right)  \\
\end{align*}

Or, 

\begin{align*}
& \sum_{c \in C} \sum_{m \in O} \frac{1}{2} \phi \left( \mu_{c,m} \right)  * \left[ \sum_{n \in O \neq m} \phi \left( \mu_{c,n} \right) \right]   \\
+ & \sum_{a \in C} \sum_{b \in C \neq a} \rho_{a.b}\sum_{o \in O}  \phi \left( \mu_{a,o} \right) \phi \left( \mu_{b,o} \right)  \\
+ & \sum_{a \in C} \sum_{b \in C \neq a} \frac{\rho_{a.b}}{2} \sum_{m \in O}  \left( \phi \left( \mu_{a,m} \right) * \left[ \sum_{n \in O \neq m} \phi \left( \mu_{b,n} \right) \right] \right)\\
\end{align*}

The expression can be simplified with the helper functions \ref{DefF} and \ref{DefG}. Applying the helper functions where they are directly applicable, the expression turns into

\begin{align*}
& \sum_{c \in C} \sum_{m \in O} \frac{1}{2} \phi \left( \mu_{c,m} \right)  * \left[ \F_T(c)  - \phi \left( \mu_{c,m} \right)\right]   \\
+ & \sum_{a \in C} \sum_{b \in C \neq a} \rho_{a.b} * \G_T(a,b)  \\
+ & \sum_{a \in C} \sum_{b \in C \neq a} \frac{\rho_{a.b}}{2} \sum_{m \in O}  \phi \left( \mu_{a,m} \right) * \left[ \F_T(b) -  \phi \left( \mu_{b,n} \right)\right] \\
\end{align*}

These can be further simplified 

\begin{align*}
& \sum_{c \in C} \sum_{m \in O} \frac{1}{2} \left[ \phi \left( \mu_{c,m} \right)  \F_T(c)  - \phi^2\left( \mu_{c,m} \right) \right]  \\
+ & \sum_{a \in C} \sum_{b \in C \neq a} \rho_{a.b} * \G_T(a,b)  \\
+ & \sum_{a \in C} \sum_{b \in C \neq a} \frac{\rho_{a.b}}{2} \sum_{m \in O} \left[  \phi \left( \mu_{a,m} \right)  \F_T(b) -   \phi \left( \mu_{a,m} \right) \phi \left( \mu_{b,n} \right)\right] \\
\end{align*}

The helper functions are now present again

\begin{align*}
& \sum_{c \in C} \frac{1}{2} \left[  \F_T(c)^2  - \G_T \left( c,c \right) \right]  \\
+ & \sum_{a \in C} \sum_{b \in C \neq a} \rho_{a.b} * \G_T(a,b)  \\
+ & \sum_{a \in C} \sum_{b \in C \neq a} \frac{\rho_{a.b}}{2} \left[ \F_T(a) \F_T(b) -  \G_T(a,b) \right] \\
\end{align*}

The second and third terms can be combined, yielding 

\begin{align*}
\sum_{c \in C} \frac{1}{2} \left[  \F_T(c)^2  - \G_T \left( c,c \right) \right] 
+ \sum_{a \in C} \sum_{b \in C \neq a} \frac{\rho_{a.b}}{2} \left[ \F_T(a) \F_T(b) +  \G_T(a,b) \right] 
\end{align*}

\begin{align*}
= \frac{1}{2} \left( \sum_{c \in C} \left[ \F_T(c)^2  - \G_T \left( c,c \right) \right] 
+ \sum_{a \in C} \sum_{b \in C \neq a} \rho_{a.b} \left[ \F_T(a) \F_T(b) +  \G_T(a,b) \right] \right)
\end{align*}

This expression has convenient symmetry. The left side has no $\rho$ term, but if it did it would be one, since $\rho_{c,c} =1$. Using the helper function $\HFunc_T(a,b)$ defined by Equation \ref{DefH}, the expression can then be rewritten as 

\begin{equation} \label{CovarianceContributionToVariance}
\frac{1}{2} \sum_{a \in C} \sum_{b \in C} \rho_{a.b} \HFunc_T(a,b) 
\end{equation}

Equation \ref{Final Variance} combines this and Equation \ref{FullVarianceTerm} to describe the full variance, $\sigma_T^2$

\subsection{Opposing teams' totals} \label{apdx.sigmaM }

By Assumptions \ref{Assumption1} and \ref{Assumption2}, the fantasy point totals of opponents are identical Normal distributions. Therefore they can be described by their shared mean and variance, $\mu_M$ and $\sigma^2_M$.

$\mu_M$ must be dependent on the number of points team $t$ scores, henceforth dubbed $Z_t$, so that the total expected value of fantasy points awarded remains constant. Therefore $\mu_M$ does not take one value- it is a function of $Z_t$, alternatively denoted $\mu_M (Z_t)$. 

Similarly, $\sigma^2_M$ is a function of values of $\mu_{c,o}$. Based on Assumption \ref{Assumption4}, for the purpose of calculating the variance of opponents, values of $\mu_{c,o}$ are random and have no dependence on the choices made by the manager in question. Therefore, even given a value of $Z_t$, an exact value of $\sigma_M$ does not exist. Still,  $\E(\sigma_M)$ and $\E(\sigma_M^2)$ can be calculated

\subsubsection{Mean}

The total number of fantasy points scored by all teams is a constant equal to the number of match-ups multiplied by the number of categories. That is, 

$$
|C| * \frac{|O| * (|O| + 1) }{2}
$$

The total number of fantasy points available for opponents is 

$$
|C| * \frac{|O| * (|O| + 1) }{2} - Z_t
$$

The expected value of points for an opponent is then 

$$
\mu_M(Z_t) = \frac{|C| * \frac{|O| * (|O| + 1) }{2} - Z_t}{|O|}
$$

Or 

\begin{equation} \label{GenericManagerMean}
\mu_M(Z_t) =  \frac{|C| (|O| + 1) }{2} - \frac{Z_t}{|O|}
\end{equation}

\subsubsection{Variance}

The expected value of $\sigma_M^2$ can be estimated based on the formula for $\sigma_T^2$ and the specifications of Assumption \ref{Assumption4}. 

According to Assumption \ref{Assumption4}, all $\mu_{c,o}$ are distributed normally and independently with mean zero and standard deviation $\sigma_c$. Call $U$ a scenario for a set of $\mu_{c,o}$ values. It can then be said that $\sigma^2_M$ is conditional upon $U$. 

For a given $U$, the variance can be calculated by Equation \ref{Final Variance}. The expected value of $\sigma_T^2$ is equal to the sum of the expected values of each of its components, which can be computed as integrals across $U$'s.

To start, consider the Bernoulli variance terms. For an arbitrary value of $U$, the probability that $\mu_{c,o} = W$ is 

$$
P(\mu_{c,o} = W) = \frac{1}{\sigma_c} \phi \left( \frac{W}{\sigma_c}\right) 
$$

The expected value of $\sum_{o \in O} \Phi(\mu_{c,o}) \left( 1 - \Phi(\mu_{c,o}) \right)$ across $U$ possibilities is then 

$$
\int_{-\infty}^{\infty} \frac{1}{\sigma_c} \phi \left( \frac{W}{\sigma_c}\right) \Phi \left( W \right) \left( 1 - \Phi( W ) \right) dW 
$$

$$
= \frac{1}{\sigma_C} \left[  \left( \int_{-\infty}^{\infty} \phi \left( \frac{W}{\sigma_c}\right) \Phi \left( W \right) dW \right) - \left( \int_{-\infty}^{\infty} \phi \left( \frac{W}{\sigma_c}\right) \Phi^2 \left( W \right) dW\right)\right] 
$$

Applying a change of variables, to $X = \frac{W}{\sigma_C}$

$$
= \frac{1}{\sigma_C} \left[  \left( \int_{-\infty}^{\infty} \sigma_c \phi \left( X \right) \Phi \left( X \sigma_c \right) dX\right) - \left( \int_{-\infty}^{\infty} \sigma_c  \phi \left( X \right) \Phi^2 \left( X \sigma_c \right) dX\right)\right] 
$$

$$
=  \left( \int_{-\infty}^{\infty} \phi \left( X \right) \Phi \left( X \sigma_c \right) dX\right) - \left( \int_{-\infty}^{\infty} \phi \left( X \right) \Phi^2 \left( X \sigma_c \right) dX\right)
$$

By Owen's integral 10,010.8 the left side is $\Phi(0) = \frac{1}{2}$ (Owen, 1980). By Owen's integral 2,0n0 the right side is 

$$
\left[ \pi - \cos^{-1} \left( \frac{\sigma_c^2}{1 + \sigma_c^2} \right) \right] \frac{1}{2 \pi}
$$

$$
=  \frac{1}{2} - \frac{\cos^{-1} \left( \frac{\sigma_c^2}{1 + \sigma_c^2} \right) }{2 \pi} 
$$

So the full expression is 

\begin{align}
\E \left( \sum_{o \in O} \Phi(\mu_{c,o}) \left( 1 - \Phi(\mu_{c,o}) \right) \right) = & \frac{1}{2} -  \left[ \frac{1}{2} - \frac{\cos^{-1} \left( \frac{\sigma_c^2}{1 + \sigma_c^2} \right) }{2 \pi}  \right] \nonumber \\
= & \frac{\cos^{-1} \left( \frac{\sigma_c^2}{1 + \sigma_c^2} \right) }{2 \pi}  \label{EPhis}
\end{align}

By assumption each component of the variance calculation from equation \ref{CovarianceContributionToVariance} is independent. The variance represented by their sum is therefore the sum of their individual variance components, conditional on a $U$ scenario. The expected value of the variance is then the sum of the expected values of the individual variance components across $U$ scenarios.

To calculate the expected values of the individual variance terms, it is necessary to calculate the expected values of the $\HFunc_T$ function for generic opponents, which will be called $\HFunc_M$. To do that, it is helpful to compute the expected values of the helper functions $\G_T$ and $\F_T$ for a generic opponent. Call them $\G_M$ and $\F_M$. 

First, consider the case when category $a$ is different from category $b$. 

With $a \neq b$, the expected value of $\F(a) \F(b)$ is the product of their expected values, since the distributions for $a$ and $b$ are independent by assumption. 

The expected value of $\F(c)$ is the sum of the expected values of its individual components. The expected value of $ \phi \left( \mu_{c,o} \right)$ is

\begin{align*}
E \left( \phi \left( \mu_{c,o} \right) \right) & = \int_{-\infty}^{\infty} \frac{1}{\sigma_c} \phi \left( \frac{W}{\sigma_C}\right) \phi \left( W \right) dW \\
& = \frac{1}{\sigma_c} \int_{-\infty}^{\infty} \phi \left( \frac{W}{\sigma_C}\right) \phi \left( W \right) dW
\end{align*}

By Owen's formula 110, this evaluates to 

\begin{align}
E \left( \phi \left( \mu_{c,o} \right) \right) & = \frac{1}{\sigma_c} \frac{1}{\sqrt{1 + \frac{1}{\sigma_c^2}}} \phi \left( 0 \right) \nonumber \\
& = \frac{1}{\sqrt{\sigma_c^2 + 1}} \phi \left( 0 \right) \label{Ephi}
\end{align}

The $F$ function has $|O|$ of these terms added together. It can then be said that 

$$
\E \left( \F_M(c) \right)= \frac{|O|}{\sqrt{\sigma_c^2 + 1}} \phi \left( 0 \right) = \frac{|O|}{ \sqrt{ 2 \pi \left( \sigma_c^2 + 1 \right) } }
$$

Based on the independence argument, $\F_M \left( a \right)$ and $ \F_M \left(b \right)$ can be multiplied together for their expectation. So 

\begin{equation} \label{EFAFB}
\E \left( \F_M \left( a \right) \F_M \left(b \right) \right) = \frac{|O|^2}{2 \pi  \sqrt{\left( \sigma_a^2 + 1 \right) \left( \sigma_b^2 + 1 \right)}}
\end{equation}

Similar reasoning can be applied to $\G(a,b)$. So long as $a \neq b$, $\mu_{a,o}$ and $\mu_{b,o}$ are independent. Therefore, $\phi \left( \mu_{a,o} \right)$ and $\phi \left( \mu_{b,o} \right)$ are independent, and the expectation of their product is the product of their expectations. Therefore it can be said that

\begin{align}
\E \left( \G_M(a,b) \right) = & |O| * \frac{1}{\sqrt{\sigma_a^2 + 1}} \phi \left( 0 \right) * \frac{1}{\sqrt{\sigma_b^2 + 1}} \phi \left( 0 \right) \nonumber \\
= & \frac{|O|}{2 \pi  \sqrt{\left( \sigma_a^2 + 1 \right) \left( \sigma_b^2 + 1 \right) } } \label{EGMAB}
\end{align}

Now the value of $H_M(a,b)$ can be described when $a \neq b$. Combining Equations \ref{EFAFB} and \ref{EGMAB}, it is 

\begin{align*}
\HFunc_M(a,b) = &
\left\{
\begin{array}{ll}
      \E \left( \F_M \left( a \right) \F_M \left(b \right) \right) + \E \left( \G_M(a,b) \right) & a \neq b\\
\end{array} 
\right. \\
= & \left\{
\begin{array}{ll}
      \frac{|O|^2}{2 \pi \sqrt{ \left( \sigma_a^2 + 1 \right) \left( \sigma_b^2 + 1 \right)}} +  \frac{|O|}{2 \pi  \sqrt{\left( \sigma_a^2 + 1 \right) \left( \sigma_b^2 + 1 \right) } } & a \neq b\\
\end{array} 
\right. 
\end{align*}

Or 

\begin{equation} \label{HMAB}
\HFunc_M(a,b) = \left\{
\begin{array}{ll}
      \frac{|O| \left(|O| + 1 \right) }{2 \pi \sqrt{ \left( \sigma_a^2 + 1 \right) \left( \sigma_b^2 + 1 \right)}} & a \neq b\\
\end{array} 
\right. 
\end{equation}

When $a = b$, the argument by independence cannot be used. The expected values must be calculated explicitly. 

For $\F_T(c)^2$, it is

\begin{equation} \label{EFC2}
\E \left( \F_T(c)^2 \right) = \sum_{m \in O} \sum_{n \in O} \E \left( \phi \left( \mu_{c,m} \right) \phi \left( \mu_{c,n} \right) \right)
\end{equation}

For $\G_T(c,c)$, it is

\begin{equation} \label{EGCC}
\E \left( \G_M(c,c) \right) = \sum_{o \in O} \E \left( \phi \left( \mu_{c,o} \right)^2 \right)
\end{equation}

Fortunately, these do not need to be computed because they simplify in the $H_M$ expression. Plugging Equations \ref{EFC2} and \ref{EGCC} into the definition of $H_T$ for $a =b$ yields 

\begin{align*}
\HFunc_M(a,b) = &
\left\{
\begin{array}{ll}
      \E \left( \F_M \left( c \right)^2 \right) + \E \left( \G_M(c,c) \right) & a = b = c\\
\end{array} 
\right. \\
= & \left\{
\begin{array}{ll}
    \sum_{m \in O} \sum_{n \in O} \E \left( \phi \left( \mu_{c,m} \right) \phi \left( \mu_{c,n} \right) \right) - \sum_{o \in O} \E \left( \phi \left( \mu_{c,o} \right)^2 \right) & a = b = c\\
\end{array} 
\right. \\
= & \left\{
\begin{array}{ll}
    \sum_{m \in O} \sum_{n \in O \neq m} \E \left( \phi \left( \mu_{c,m} \right) \phi \left( \mu_{c,n} \right) \right)  & a = b = c\\
\end{array} 
\right. 
\end{align*}

The result is $|O| ( |O| - 1 ) $ terms of products of $\phi$ represented by different opponents. The expected value of one $\phi$ term is $\frac{1}{\sqrt{\sigma_c^2 + 1}} \phi \left( 0 \right)$ by Equation \ref{Ephi}. By independence, the expected value of the product of two is the square of that expectation, $\frac{1}{\sigma_c^2 + 1} \phi \left( 0 \right)^2$ . Therefore

\begin{align}
\HFunc_M(a,b) = &
\left\{
\begin{array}{ll}
      |O| ( |O| - 1 )  * \frac{1}{\sigma_c^2 + 1} \phi \left( 0 \right)^2 & a = b = c\\
\end{array} 
\right. \nonumber \\
= & \left\{
\begin{array}{ll}
    \frac{ |O| ( |O| - 1 )}{ 2 \pi \left( \sigma_c^2 + 1 \right) }  & a = b = c\\
\end{array} 
\right. \label{HMC}
\end{align}

Between Equations \ref{HMAB} and \ref{HMC}, the full $\HFunc_M$ function can be written as per Equation \ref{GM}. Combining that with the result from Equation \ref{EPhis}, the total variance is then reflected by Equation \ref{Generic Variance}.

Note that since variance is the sum of several independent terms, the central limit theorem applies and it is roughly a Normal distribution. As a large Normal distribution which is always far above $0$, Lemma \ref{Approx 3} dictates that its square root is also approximately a Normal distribution with mean equal to the square root of the mean of the variance. This means that the square root of the expected variance can be used as an approximation for the expected standard deviation, justifying Equation \ref{SDVarEq}

\begin{equation} \label{SDVarEq}
\E(\sigma_M) = \sqrt{ \E(\sigma^2_M) }
\end{equation}

\subsection{Fantasy point total required to win}

Call the fantasy point total required to win $Z_R$. It is equal to the highest fantasy point total among opponents. It can also be decomposed into two components; the average total for an opponent $\mu_M$, and the highest deviation above $\mu_M$ among opponents, dubbed $L$. Numerically,

\begin{equation} \label{LEq}
Z_R = \mu_M + L
\end{equation}

Since $\mu_M$ is already known, attention can shift to the properties of $L$. By Assumption \ref{Assumption3}, $L$ is a Normal distribution. It is then important to know the mean and variance of $L$, $\mu_L$ and $\sigma^2_L$

\subsubsection{Expected value}

Based on Lemma \ref{Approx2}, The expected value of the maximum of $N$ Normal variables is approximately 

$$
\sigma \MEV(N)
$$

In this case, $\sigma$ is conditional upon $U$ and $\MEV(N)$ is a constant. Therefore the expected value of the maximum is the expected value of $\sigma$ times the $\MEV$ constant. By equation \ref{SDVarEq}, that means it can be represented by Equation \ref{MUL}

\subsubsection{Variance}

Based on Lemma \ref{Approx2}, the variance is roughly 

$$
\sigma^2 \MVAR(N)
$$

Again, the expected value of this quantity is the expected value of the variance times $\MVAR$, which in this case is Equation \ref{SigmaL}

\subsection{Differential between team $t$ and target}

Given that team $t$ scores $Z_t$, the differential is by definition $D = Z_t - Z_R$. Substituting in Equation \ref{LEq}, that can be rewritten to

$$
D = Z_t - \mu_M(Z_t) - L
$$

Using Equation \ref{GenericManagerMean} to define $\mu_M$ explicitly yields

\begin{align*}
D = & Z_t - \frac{|C| (|O| + 1) }{2} + \frac{Z_t}{|O|} - L
\end{align*}

This can be further simplified, to make a more convenient description of $D$

\begin{equation} \label{DefD}
D = Z_t \frac{|O| + 1}{|O|}- \frac{|C| (|O| + 1) }{2} - L
\end{equation}

The quantities of interest are the mean and variance of $D$, $\mu_D$ and $\sigma_D^2$

\subsubsection{Mean}

The mean value of $Z_t$ is $\mu_T$, and the mean of $L$ is $\mu_L$. Applying those values to Equation \ref{DefD}, the result is Equation \ref{MUD}

\subsubsection{Variance}

The variance of $Z_t$ is $\sigma^2_T$ and the variance of $L$ is defined as $\sigma^2_L$. Applying those to Equation \ref{DefD} yields Equation \ref{SUD}

\subsection{Victory probability}

By Assumptions \ref{Assumption1} and \ref{Assumption3}, it is apparent that the distribution of the differential is a Normal distribution. It has mean $\mu_D$ and variance $\sigma_D^2$. From this, Equation \ref{VProb} follows as a description of the victory probability

\section{Gradient of $V$} \label{apdx.Grad}

$V$ is a function of $\mu_D$ and $\sigma_D$. To get its gradient, the gradients of its components can be evaluated then combined based on the definition of $V$

\subsection{Gradient of $\mu_D$} \label{mudgradjust}

The only variable component of $\mu_D$ as described by Equation \ref{MUD} is $\mu_T$. Therefore its gradient is just $\frac{|O| + 1}{|O|}$ times the gradient of $\mu_T$. 

Relative to $\mu_{c,o}$ the gradient of $\mu_T$ defined by Equation \ref{Mean} is simply the PDF of the corresponding Normal distribution. That is, 

\begin{equation} \label{MUPGrad}
\nabla_{c,o} (\mu_T) = \phi(\mu_{c,o})
\end{equation}

Multiplying the value from Equation \ref{MUPGrad} by $\frac{|O| + 1}{|O|}$ yields Equation \ref{mudGrad}

\subsection{Gradient of $\sigma_D$}

The gradient of $\sigma_D$ is 

\begin{align} 
\nabla_{c,o} \left( \sigma_D \right) = & \nabla_{c,o} \sqrt{\sigma_T^2 + \sigma_T^2} \nonumber \\
                                     = &\frac{1}{2 \sqrt{\sigma_T^2 + \sigma_T^2}}\nabla_{c,o} \left( \sigma_T^2 + \sigma_T^2 \right) \nonumber \\ 
                                     = &\frac{1}{2 \sigma_D}\nabla_{c,o} \left( \sigma_T^2 \right) \label{sigmaDGrad}
\end{align}

It is then necessary to compute the gradient of $\sigma_T^2$ as described by Equation \ref{SigPGrad}. Deriving it is a somewhat involved calculation

\subsubsection{Gradient of the variance terms}

Relative to $\mu_{c,o}$ the gradient of the $\Phi$ terms is

$$
\nabla_{c,o} \left[ \Phi(\mu_{c,o}) \left( 1 - \Phi(\mu_{c,o}) \right) \right]
$$

$$
= \nabla_{c,o} \left[ \Phi(\mu_{c,o}) - \Phi(\mu_{c,o})^2 \right] 
$$

$$
= \phi(\mu_{c,o}) - 2 \Phi(\mu_{c,o}) \phi(\mu_{c,o}) 
$$

\subsubsection{Gradient of the covariance terms}

The gradient of the $\HFunc_T$ terms are dependent on the gradients of $\F$ and $\G$.

With respect to a single $c$ and opponent $o$, the gradient of $F_T(c)$ is

$$
\nabla \phi(\mu_{c,o})
$$

This can be computed by the reverse of Owen's Integral 11 (Owen, 1980). It is

$$
- \mu_{c,o} \phi(\mu_{c,o})
$$

The gradient of $F_T(c)^2$ is 

\begin{align*}
\nabla_{c,o} F_T(c)^2 = & 2 F_T(c) \nabla_{c,o} F_T(c) \\
= & - 2 F_T(c) \mu_{c,o} \phi(\mu_{c,o})
\end{align*}

The gradient of $F_T(b)$ is of course zero relative to $\mu_{c,o}$. So the gradient of $F_T(a)F_T(b)$ relative to $\mu_{c,o}$ is 

\begin{equation} \label{FGrad}
\nabla_{c,o} F_T(a)F_T(b) = \left\{
\begin{array}{ll}
      - 2 * F_T(c) * \mu_{c,o} * \phi(\mu_{c,o}) & a = c, b = c \\
      - \mu_{c,o} \phi(\mu_{c,o}) F_T(b) & a = c, b \neq c\\
       0 & a \neq c, b \neq c\\
\end{array} 
\right. 
\end{equation}

The gradient of $G_T(b, c)$ with respect to a single $\mu_{c,o}$ is 

$$
\nabla_{c,o} \left( G_T(b, c) \right)= \nabla_{c,o} \phi(\mu_{c,o}) * \phi(\mu_{b,o})
$$

Which is the same as the derivative for $F_T(c)$, except with the extra factor of $\phi(\mu_{b,o})$. That is, 

$$
\nabla_{c,o} \left( G_T(b, c) \right) = - \mu_{c,o} \phi(\mu_{c,o}) \phi(\mu_{b,o})
$$

The gradient of $G_T(c,c)$ is 

\begin{align}
\nabla_{c,o} \left( G_T(c,c) \right) =  & \nabla_{c,o} \left(  \phi(\mu_{c,o})^2 \right) \\
= & 2  \phi(\mu_{c,o}) \nabla \phi(\mu_{c,o}) \\
= & - 2  \mu_{c,o} \phi(\mu_{c,o})^2
\end{align}

So the gradient of $G$ is 

\begin{equation} \label{GGrad}
\nabla_{c,o} \left( G_T(a,b) \right) = \left\{
\begin{array}{ll}
      - 2 \mu_{c,o} \phi(\mu_{c,o})^2 & a = c, b = c \\
      - \mu_{c,o} \phi(\mu_{c,o}) \phi(\mu_{b,o}) & a = c, b \neq c\\
       0 & a \neq c, b \neq c\\
\end{array} 
\right. 
\end{equation}

Combining Equations \ref{FGrad} and \ref{GGrad}, the gradient of $H_T(a,b)$ is 

\[ 
\nabla_{c,o} \left( \HFunc_T(a,b) \right) = \left\{
\begin{array}{ll}
      - 2 F_T(c) \mu_{c,o} \phi(\mu_{c,o}) +  2 \mu_{c,o} \phi(\mu_{c,o})^2 & a = c, b = c \\
      - \mu_{c,o} \phi(\mu_{c,o}) F_T(b) - \mu_{c,o} \phi(\mu_{c,o}) \phi(\mu_{b,o}) & a = c, b \neq c\\
       0 & a \neq c, b \neq c\\
\end{array} 
\right. 
\]

The gradient of the full covariance term is 

$$
\frac{1}{2} \sum_{a \in C} \sum_{b \in C \neq a} \rho_{a.b} \nabla_{c,o} \left( \HFunc_T(a,b) \right)
$$

The terms for which neither $a$ nor $b$ are $c$ can be ignored because their gradients are zero. That leaves one term for $a = b = c$, and double-counted terms for pairs of $c$ and all other categories. Written out, this becomes

$$
\frac{1}{2} \left[ \left( 2 \sum_{b \in C \neq c} \rho_{b,c} \left[ - \mu_{c,o} \phi(\mu_{c,o}) \phi(\mu_{b,o}) - \mu_{c,o} \phi(\mu_{c,o}) F_T(b)\right] \right) - 2 F_T(c) \mu_{c,o} \phi(\mu_{c,o}) + 2 \mu_{c,o} \phi(\mu_{c,o})^2 \right]
$$

Which can be simplified to 

$$
\mu_{c,o} \phi(\mu_{c,o}) \left[ \left( \sum_{b \in C \neq c} \rho_{b,c}  \left[ - \phi(\mu_{b,o}) -  F_T(b)\right] \right) +  \left(\phi(\mu_{c,o}) - F_T(c) \right) \right]
$$

\subsubsection{Final tally} \label{sigpgradjust}

Combining the results from the variance and covariance terms, the full gradient of $ \sigma_T^2$ relative to $c$, $o$ is then represented by Equation \ref{SigPGrad}

\subsection{Gradient of $V$}

The gradient of $V$ as defined by Equation \ref{VProb} is 

$$
 \nabla_{c,o} (V) =  \nabla_{c,o} \Phi \left(\frac{\mu_D}{\sigma_D} \right)
$$

By the chain rule, this is 

$$
 \nabla_{c,o} (V) = \phi \left(\frac{\mu_D}{\sigma_D} \right) \nabla_{c,o} \left[ \frac{\mu_D}{\sigma_D}\right]
$$

Invoking the quotient rule, this is

$$
 \nabla_{c,o} (V) = \phi \left(\frac{\mu_D}{\sigma_D} \right)  \left( \frac{ \sigma_D * \nabla_{c,o} \left( \mu_D \right) - \mu_D * \nabla_{c,o} \left( \sigma_D
 \right) }{\sigma^2_D} \right)
$$

Equation \ref{sigmaDGrad} can be plugged in, transforming the equation into

\begin{align*}
  \nabla_{c,o} (V) = & \phi \left(\frac{\mu_D}{\sigma_D} \right)  \left( \frac{ \sigma_D * \nabla_{c,o} \left( \mu_D \right) - \mu_D * \frac{\nabla_{c,o} \left( \sigma_T^2
 \right)}{2 \sigma_D} }{\sigma^2_D} \right) \\
= & \phi \left(\frac{\mu_D}{\sigma_D} \right)  \left( \frac{ \sigma_D^2 * \nabla_{c,o} \left( \mu_D \right) - \frac{\mu_D}{2} * \nabla_{c,o} \left( \sigma_T^2
 \right) }{\sigma^3_D} \right)
\end{align*}

Thus justifying Equation \ref{Grad}. The necessary subgradients are described by Equations \ref{SigPGrad} and \ref{mudGrad}, justified in Appendices \ref{mudgradjust} and \ref{sigpgradjust}

\section{Lemmas} \label{apdx.Lemmas}

\subsection{Lemma 1} \label{Lemma1}

The PDF of the standard bivariate normal is 

$$
\frac{1}{2 \pi \sqrt{1 - \rho^2}} e^{\frac{- \left(x^2 - 2 \rho x y + y^2 \right) }{2\left(1 - \rho^2 \right) }}
$$

This can be separated into 

$$
\frac{1}{2 \pi \sqrt{1 - \rho^2}} \left[ e^{\frac{-\left(x^2 \right) }{2\left(1 - \rho^2 \right) }} * e^{\frac{-\left(y^2 \right) }{2\left(1 - \rho^2 \right) }} * e^{\frac{\left( 2 \rho x y \right) }{2\left(1 - \rho^2 \right) }} \right]
$$

When $\rho$ is not large, it is possible to invoke the approximation that $1 - \rho^2 \approx 1$.  The expression can then be reduced to 

\begin{align*}
& \frac{1}{2 \pi} \left[ e^{\frac{-\left(x^2 \right) }{2}} * e^{\frac{-\left(y^2 \right) }{2 }} * e^{\frac{\left( 2 \rho x y \right) }{2 }} \right] \\
= & \frac{1}{\sqrt{2 \pi}} e^{\frac{-x^2}{2}} * \frac{1}{\sqrt{2 \pi}} e^{\frac{-y^2 }{2 }} * e^{ \rho x y }  \\
= & \Phi(x) * \Phi(y) * e^{\rho x y } 
\end{align*}

The first order Taylor series for $e^z$ approximates this as

\begin{align*}
& \phi(x) * \phi(y) * \left( 1 +  \rho x y \right) \\
= & \phi(x) \phi(y) +   \rho \phi(x) \phi(y) x y
\end{align*}

Calculating the integral of this PDF up to $x = a$ and $y = b$ will yield the corresponding CDF, which is the quantity of interest. First, integrating by $x$

$$
BvN(a,b,\rho) = \int_{x=-\infty}^{x =a} \phi(x) \phi(y) +   \rho \phi(x) \phi(y) x y
$$

According to Owen's integral table, $\int x \phi (x) = -\phi(x)$ (Owen, 1980). So this is

$$
BvN(a,b,\rho) = \Phi(a) \phi(y) + \rho * \left( - \phi(a) \right)  * \phi(y)  y 
$$

Doing the same for $y$ yields 

\begin{align*}
BvN(a,b,\rho) = & \int_{y=-\infty}^{y = b} \Phi(a) \phi(y) + \rho * \left( - \phi(a) \right)  * \phi(y)  y   \\
= & \Phi(a) \Phi(b) + \rho \left( - \phi(a) \right) * \left( - \phi(b) \right)  \\
= & \Phi(a) \Phi(b) + \rho \phi(a) \phi(b) 
\end{align*}

Arriving at the approximation described in the Lemma

\subsection{Lemma 2} \label{apdx.Lemmas2}

An explicit formula for the expected value and variance of the maximum of $N$ standard identical Normal distributions is not available for arbitrary values of $N$. Fortunately, values for $N \leq 20$ have been estimated precisely by previous work (Teichroew, 1956). 

Teichroew provides tables of expected values of order statistics of $N$ independent standard Normals, and expected values of products of those order statistics. The maximum is equivalent to Teichroew's first order statistic. This allows the expected value of the maximum to be transcribed directly from his tables. The expected value of the maximum squared can also be found in Teichroew's tables, as the product of the first order statistic and itself. Variance can then be computed by applying the fact that variance is equal to $\E(X^2) - \E(X)^2$. In Table \ref{FullMaxTable}, the first two columns are transcribed from Teichroew's tables, and the third is computed as the second minus the first squared. The first and third columns are $\MEV$ and $\MVAR$ respectively.

\begin{table}[!ht] 
    \centering
    \begin{tabular}{|l|l|l|l|}
    \hline
        N & $\E(X)$ & $\E(X^2)$ & $\Var(X)$ \\ \hline
        1 & 0 & 1 & 1 \\ \hline
        2 & 0.564189584 & 1 & 0.681690114 \\ \hline
        3 & 0.846284375 & 1.275664448 & 0.559467204 \\ \hline
        4 & 1.029375373 & 1.551328895 & 0.491715237 \\ \hline
        5 & 1.162964474 & 1.800020436 & 0.447534069 \\ \hline
        6 & 1.267206361 & 2.021739069 & 0.415927109 \\ \hline
        7 & 1.352178376 & 2.220304137 & 0.391917777 \\ \hline
        8 & 1.423600306 & 2.399534975 & 0.372897143 \\ \hline
        9 & 1.485013162 & 2.562617418 & 0.357353326 \\ \hline
        10 & 1.538752731 & 2.71210379 & 0.344343823 \\ \hline
        11 & 1.586436352 & 2.850027741 & 0.333247443 \\ \hline
        12 & 1.62922764 & 2.97801909 & 0.323636387 \\ \hline
        13 & 1.667990177 & 3.097396615 & 0.315205384 \\ \hline
        14 & 1.703381554 & 3.209238821 & 0.307730102 \\ \hline
        15 & 1.735913445 & 3.314427059 & 0.30103157 \\ \hline
        16 & 1.766991393 & 3.413735409 & 0.291476826 \\ \hline
        17 & 1.793941081 & 3.507760835 & 0.289536233 \\ \hline
        18 & 1.820031879 & 3.59704617 & 0.28453013 \\ \hline
        19 & 1.844481512 & 3.682047852 & 0.279935805 \\ \hline
        20 & 1.86747506 & 3.763159715 & 0.275696616 \\ \hline
    \end{tabular}
    \caption{Full table of data transcribed from Teichroew, and the computed variance. The first colum is $\MEV(N)$ and the third column is $\MVAR(N)$}
    \label{FullMaxTable}
\end{table}

This is sufficient for the algorithm to handle leagues up to $|O| = 20$. If values for larger $N$ are needed, they could be estimated either by applying heuristics or similar numerical methods to Teichroew's

\subsection{Lemma 3}

Consider $X$ to be a Normal distribution with mean $\mu$ and variance $\sigma^2$. The CDF of $X$ at $x$ is 

$$
\Phi(\frac{x - \mu}{\sigma})
$$

Say $Y = \sqrt{X}$, or equivalently. $X = Y^2$. In that case, if $X \leq y^2$, then $Y^2 \leq y^2$. So long as $Y$ is positive, which by assumption it is, that is equivalent to $Y < y$. Therefore, the probability that $Y < y$ is the same as the probability that $X \leq y^2$, and

$$
P(Y \leq y) = \Phi(\frac{y^2 - \mu}{\sigma})
$$

The assumption of the Lemma is that $X$ is near its mean. Therefore in the square root basis, $y$ is near the square root of the mean. Accordingly, $y$ can be redefined as $\sqrt{\mu} + h$, where $h$ is an arbitrarily small factor. The CDF becomes 

\begin{align*}
P(Y \leq y) = & \Phi \left( \frac{\left(\sqrt{\mu} + h\right)^2 - \mu}{\sigma} \right) \\ = &
\Phi \left( \frac{\mu + 2h\sqrt{\mu} + h^2 - \mu}{\sigma} \right) \\ = &\Phi \left( \frac{2h\sqrt{\mu} + h^2}{\sigma} \right)
\end{align*}

$h$ is small, the $h^2$ term can be dropped. Also, the remaining $h$ term can be rewritten to $y - \sqrt{\mu} $, leading to

\begin{align*}
P(Y \leq y) =  & \Phi \left( \frac{2 h \sqrt{\mu} }{\sigma} \right) \\
= & \Phi \left( \frac{2 \left( y - \sqrt{\mu} \right)\sqrt{\mu} }{\sigma} \right) \\ = &  
\Phi \left( \frac{y - \sqrt{\mu} }{ \frac{\sigma}{2 \sqrt{\mu}} } \right)
\end{align*}

This is recognizable as the CDF of a Normal distribution with mean $\sqrt{\mu}$ and standard deviation $ \frac{\sigma}{2 \sqrt{\mu}}$

\end{appendices}

\bibliographystyle{agsm}

\end{document}